\documentstyle[prb,aps,epsfig]{revtex}
\begin{document}
\flushbottom
\draft          % \draft command makes pacs numbers print
\wideabs{       % makes abstract and title in one column
%%%%%%%%%%%%%%%%%%%%%%%%%%%%%%%%%%%%%%%%%%%%%%%%%%%%%%%%%%%
\title{Enhancement of Rabi Splitting in a Microcavity with an Embedded Superlattice}
\author{J. H. Dickerson* and E. E. Mendez}
\address{Department of Physics and Astronomy, State University of New York at Stony Brook, Stony Brook, NY 11794-3800}
\author{A. A. Allerman}
\address{Sandia National Laboratory, Albuquerque, NM 87185}
\author{S. Manotas, F. Agull\'{o}-Rueda, and C. Pecharrom\'{a}n}
\address{Instituto de Ciencia de Materiales, CSIC, Cantoblanco, 28049 Madrid, Spain}
\date{\today}
\maketitle
\begin{abstract}
We have observed a large coupling between the excitonic and photonic modes of an AlAs/AlGaAs microcavity filled with an 84-\({\rm {\AA}}\)/20\({\rm {\AA}}\) GaAs/AlGaAs superlattice.  Reflectivity measurements on the coupled cavity-superlattice system in the presence of a moderate electric field yielded a Rabi splitting of 9.5 meV at T = 238 K.  This splitting is almost 50\(\% \) larger than that found in comparable microcavities with quantum wells placed at the antinodes only.  We explain the enhancement by the larger density of optical absorbers in the superlattice, combined with the quasi-two-dimensional binding energy of field-localized excitons.
\end{abstract}
\vspace{-.4cm}
}  % end of wideabs
The ability of semiconductor multilayers to mimic atomic cavities has allowed the observation of 
phenomena that until recently seemed reserved to atomic physics.  Among these, the best studied 
so far is the coupling of a two-level atom and the electromagnetic modes of a cavity enclosing it,
which results in the so-called Rabi splitting when there is a resonance between one of those modes
and the energy difference between the atomic levels. ~\cite{Zhu:90}  Analogously, in a semiconductor
microcavity formed by two dielectric mirrors and a quantum well (or a small set of quantum wells)
in the cavity between them, the excitonic mode of the quantum well is coupled to a photonic mode of
the cavity when the characteristic energies of the two modes, h\(\nu _{exe}\) and h\(\nu _{cav}\),
coincide. ~\cite{Weisbuch:92,Savona:95,Stanley:96}

If strong enough, this coupling is manifested as a splitting into two of the optical features
associated with that energy, for instance, a minimum in the reflectivity spectrum or a peak in the
absorption or luminescence spectrum.  In atomic systems the Rabi splitting is seen easily at room 
temperature, but in semiconductors the splitting is much less apparent because of a smaller 
splitting-to-linewidth ratio and a considerable decrease with increasing temperature.

The magnitude of the splitting is proportional to the overlap between the exciton's wavefunction
and the photon's electric field and to the square root of the product of oscillator strength and 
linear density of excitons. ~\cite{Houdre:94}   Therefore, it is not surprising that a considerable
effort has been made to increase these factors.  To this end, more than one quantum well has been
placed at the antinodes of the mirror-defined cavity, and various material systems have been used.
In microcavities made out of III-V semiconductors, such as GaAs, AlAs, and InAs, typical Rabi
splittings are 6 to 7 meV at low temperature and less than 5 meV at room 
temperature. ~\cite{Houdre:94,Tao:98,Pratt:98,Houdre:94b}  Because of their larger exciton
binding energy (and corresponding larger oscillator strength) CdTe-based microcavities have shown
splittings as large as 23 meV at 4 K and 17 meV at 300 K. ~\cite{Dang:98,Boeuf:00}

The Rabi splitting is often limited by the imperfect confinement of the photon field within the
physical boundaries of the cavity, which reduces the photon-exciton overlap.

\begin{figure}
\psfig{file=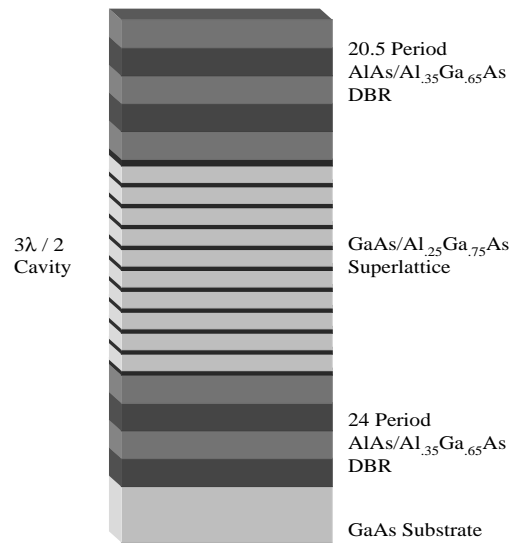,height=3.5in,width=3in}
\caption{Two Distributed Bragg reflector mirrors (DBRs), comprised of alternating \(\frac{{\lambda }}{4} \) layers of 689 \({\rm {\AA}}\) AlAs and 602 \({\rm {\AA}}\) \({\rm Al}_{{\rm .35}} {\rm Ga}_{{\rm .65}} {\rm As}\), flanking an 84 \({\rm {\AA}}\) GaAs / 20 \({\rm {\AA}}\) \({\rm Al}_{{\rm .25}} {\rm Ga}_{{\rm .75}} {\rm As}\) superlattice that defines a \(\frac{{3\lambda}}{2} \) cavity.  The DBRs were either undoped (Sample B) or doped (Sample A) p-type (1\( \times \)\(10^{18} \) of Be), for the top mirror, and n-type (1\( \times \)\(10^{18} \) of Si) for the bottom mirror.}
\end{figure}

To ameliorate that effect, selectively oxidized AlAs/GaAs mirrors, with a larger index-of-refraction
contrast, have been used. ~\cite{Nelson:96,Pratt:99}  Others have embedded quantum wells in the mirror
regions, thus increasing that overlap. ~\cite{Bloch:98}  In still another scheme the entire cavity
between the two mirrors consisted of just one material (GaAs), thus forming a "bulk" microcavity in
which the larger number of excitons per unit length partially compensated both for the smaller binding
energy of three-dimensional (3D) excitons and the spatially non-uniform coupling to the photon 
field. ~\cite{Tredicucci:95}  The Rabi splitting found experimentally for these structures was about
half of that in comparable quantum-well microcavities.

In this Communication we present a simple way of enhancing the exciton-cavity coupling in a 
quantum-well microcavity system. ~\cite{Dickerson:01}  By exploiting the localization properties 
of a superlattice under a longitudinal electric field, we have overcome the limitation in a previous
scheme brought out by the small binding energy of a 3D exciton, while keeping the benefits of a thick
active medium.  Our approach consists of filling the entire cavity with a superlattice subjected to 
an electric field, rather than placing a few isolated wells at the antinodes of the cavity.
A narrow-barrier superlattice within the active region maximizes the density of quantum wells while
the field enhances the exciton binding energy through Wannier-Stark localization of the superlattice
states.~\cite{Wannier:60,Mendez:88}  Using this scheme we have found a 9.5 meV Rabi splitting at
T = 238 K, which represents almost a \(50\% \) increase of the splitting found in a comparable optical
structure in which the field is absent and quantum wells are placed exclusively at the antinodes.

In a superlattice of period D, the electronic wavefunctions are extended, and the exciton's binding 
energy is comparable to that of a 3D semiconductor.  However, under a longitudinal (that is, parallel
to the superlattice direction) electric field \(\varepsilon \), the spatial extension of the 
wavefunctions is reduced to a length of the order of \(\frac{\Delta }{{e\varepsilon }}\), where
\(\Delta \) is the energy width of the miniband, and e is the electronic charge.  Beyond a certain
field, electrons and holes become localized into individual quantum wells, excitons become quasi
two-dimensional, and their binding energy increases drastically.  Then, it should be possible to 
achieve enhanced Rabi splitting by filling a semiconductor microcavity with a superlattice that is 
subjected to a strong enough electric field, applied either internally or externally.

The microcavities we have used to test this idea consisted of two distributed Bragg reflector mirrors
(DBRs) flanking a 30 period 84-\({\rm {\AA}}\) GaAs / 20-\({\rm {\AA}}\) \({\rm Al}_{{\rm .25}} 
{\rm Ga}_{{\rm .75}} {\rm As}\) superlattice, which formed the active region of the cavity (see Fig. 1).
The top (bottom) reflector had 20.5 (24) periods of alternating 689-\({\rm {\AA}}\) AlAs and
602-\({\rm {\AA}}\) \({\rm Al}_{{\rm .35}} {\rm Ga}_{{\rm .65}} {\rm As}\) \(\frac{\lambda }{4}\) 
layers.  In sample A, the top mirror was p-type doped (Be,1 \(\ times \) \(10^{18} \) \(cm^{-3} \)
and the bottom mirror was doped n-type (Si, 1 \(\ times \) \(10^{18} \) \(cm^{-3} \), while the cavity
was undoped, thus forming a p-i-n heterostructure.  In sample B, the entire structure was undoped. 
The number of superlattice periods was chosen so that the length of the cavity was
\(\frac{3\lambda }{2}\), where \(\lambda \) = \(\frac{\lambda _0 }{n}\).  \(\lambda _0 \) is 
the wavelength associated with the exciton binding energy \(E_{ex} \) 
(\(\lambda _0 \)= \(\frac{{ch}}{{E_{ex} }}\)); n is the active region's index of refraction. c is
the speed of light in vacuum, and h is Planck's constant.  In our case,  \(\lambda _0\) equaled 8040
\({\rm {\AA}}\), which corresponds to the e1--hh1 exciton energy for the 84
\({\rm {\AA}}\)/20 \({\rm {\AA}}\) superlattice at around liquid-nitrogen temperature.
We also prepared a quantum-well microcavity (sample C) in which the superlattice of sample A
was replaced by two uncoupled 89 \({\rm {\AA}}\) quantum wells at each of the three anti-nodes
of the \(\frac{{3\lambda }}{2}\) cavity.  The focus of the work reported here is on sample A, with
samples B and C serving as controls.

\begin{figure}
\psfig{file=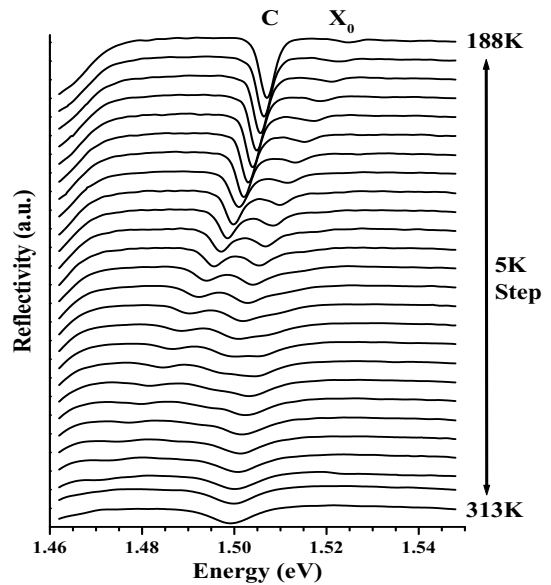,height=3.5in,width=3in}
\caption{Experimental reflectivity spectra for the doped microcavity (sample A in the text) across the 188 K to 313 K temperature range, in 5 K intervals.  For clarity, the spectra have been shifted vertically in a cascade format.  Two minima are easily observed at most temperatures, while an additional, very weak minimum can be discerned at some temperatures.}
\end{figure}

The Rabi splitting was determined primarily from reflectivity measurements, complemented by 
photoluminescence experiments.  The energy coupling between the exciton and the cavity-photon 
modes was tuned in and out of resonance by varying either the temperature of the sample or the spot
that was probed.  A slight taper of the microcavity, inherent to the epitaxial MOCVD growth process,
shifted the energy of the cavity mode along the radial direction of the 3" GaAs wafers on which the 
multilayers were grown.  The cavity mode shifted approximately 10 meV (and in one sample up to 30 meV)
from the center to the edge, while the exciton energy remained practically unchanged.  On the other 
hand, the energy of the exciton changed with temperature at a rate of 3.8 meV per 10 K, a rate more 
than five times larger than that of the cavity mode.

The reflectivity of the microcavities was measured at normal incidence, using a micro-reflectivity
spectroscopy set-up coupled to a broad-spectrum quartz tungsten halogen lamp and a filter to block 
photon energies above 1.9 eV.  The spatial and spectral resolutions of the system are better than 2
\(mu m\) and 0.20 meV, respectively.

Figure 2 shows reflectivity spectra for sample A in the temperature range 188 K--313 K in the 
1.46 eV--1.55 eV spectral region, which corresponds to the high-reflectivity region of the Bragg
reflectors.  At the lowest temperature the strongest reflectivity minimum is at 1.5078 eV, with a 
second one at 1.5238 eV and a third, barely visible dip at 1.5382 eV.  As the temperature increases, 
the latter two minima shift to lower energy faster than the former, and simultaneously the relative 
intensity of the three minima gradually changes.  Above 260 K, the highest-energy minimum is most 
pronounced and shifts the least as the temperature increases.

The energy positions of the three minima as a function of temperature are summarized in Fig. 3. 
The linear shift of these features at the two ends of the temperature range attests their origin:
at low T, the stronger, low-energy feature evidences the cavity mode (labeled C in the figure), 
whereas the two features at higher energy have excitonic character and are labeled as \(X_0 \) and 
\(X_1 \) in Fig. 3.  At high T, the situation is reversed: the stronger, high-energy minimum 
corresponds to the cavity mode and the much weaker minima at low energy are excitonic in nature. 
At intermediate temperatures, two anticrossings are observed, one between C and \(X_0 \) at T = 238 K 
and another between C and \(X_1 \) at T = 253 K, with anticrossing energies of 9.5 meV and 6.9 meV,
respectively.

We interpret the \(X_0 \) line as the electric-field localized e1--hh1 exciton of the superlattice,
which interacts with the cavity mode C and forms a mixed state at intermediate temperatures. The 
minimum 9.5 meV splitting corresponds to the maximum coupling and is a measure of the Rabi splitting. 
This interpretation is supported by temperature-dependent photoluminescence measurements, which show 
a behavior and anticrossing similar to that of \(X_0 \) and C in Fig. 3.  Room temperature, 
position-dependent reflectivity measurements on a different portion of sample A exhibited two clear
minima whose relative strength and energy separation depend on the spot probed, from which a Rabi 
splitting of 8.5 meV was inferred.

The 9.5 meV splitting is almost 50\(\% \) larger than the splitting obtained in similar 
temperature-dependence reflectivity measurements in sample C, in which the superlattice was
replaced by quantum wells at the antinodes.  This shows that the larger number of wells favors a
larger Rabi splitting, even though many are not in optimal positions in the cavity to couple
effectively to the optical field.  That the superlattice excitons are localized by the electric field
is essential to the enhancement of Rabi splitting, as evidenced by the 4.5 meV splitting obtained for 
sample B.  In sample A, the built-in field of the p-i-n structure provides substantial, if not complete
(see below), wavefunction localization for both electrons and holes.  In contrast, the cavity in
sample B is in flat-band configuration; the electrons and holes are delocalized and the exciton 
binding energy is comparable to that of bulk material.

The origin of the \(X_1 \) line in Fig. 3, the line responsible for a secondary anticrossing of
6.9 meV with the cavity mode at T = 253 K, is less clear.  One possibility is that \(X_1 \) is 
associated with the field-localized light-hole exciton (e1--lh1) of the superlattice.  The calculated
energy difference between the first heavy-hole and light-hole states in an isolated 
84-\({\rm {\AA}}\) GaAs / \({\rm Al}_{{\rm .25}} {\rm Ga}_{{\rm .75}} {\rm As}\) quantum well
is 18 meV, which is not very different from the 15 meV average separation between the \(X_0 \)
and \(X_1 \) lines when both are far from the anticrossing region with the cavity mode C. ~\cite{Info1}
Within this interpretation, the weakness of the reflectivity minima associated to \(X_1 \) (see Fig. 2)
could be explained by the smaller oscillator strength of the light-hole exciton.

\begin{figure}
\psfig{file=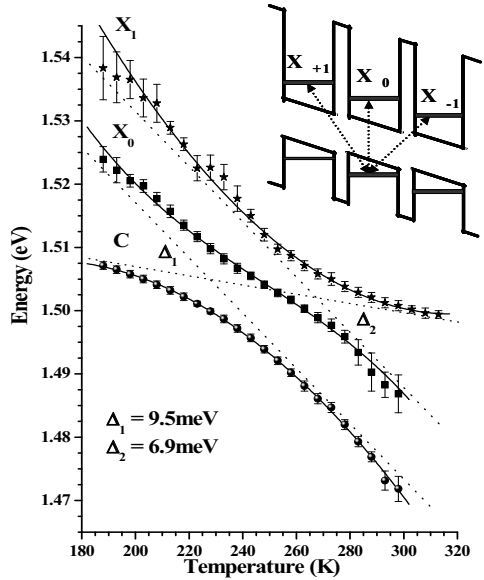,height=3.5in,width=3in}
\caption{Energies of the reflectivity minima in Fig. 2 plotted as a function of temperature.  The two rapidly changing energies correspond to 0 and +1 heavy-hole exciton transitions of the Stark ladder (\(X_0 \) and \(X_1 \), respectively) whereas the more gradually changing energy corresponds to the cavity mode, C.  The dotted lines are guides to the eye to indicate the temperature trends of the uncoupled exciton and cavity modes.}
\end{figure}

Another possibility is that the \(X_1 \) line is due to a spatially indirect exciton involving
electrons in one quantum well of the superlattice and holes in an adjacent well (see the inset 
of Fig. 3) that is, the +1 exciton involving the electron and hole Wannier-Stark ladders of the
superlattice.  Although at high electric fields (\(\varepsilon  > > \frac{\Delta }{{eD}}\)) there
is complete localization of electrons and holes to individual wells, at smaller fields the 
localization is incomplete, with the electronic wavefunctions extended beyond a single well. 
In sample A, the calculated energy widths of the electron and heavy hole minibands are 19.5 meV 
and 1.5 meV, respectively, which implies that the high-field regime is reached for
\(\varepsilon  >  > \) 20 kV/cm.  If we assume that the built-in electric field in sample A's
p-i-n configuration does not penetrate into the doped dielectric mirrors, then that field is
\( \approx \) 50 kV/cm, which is sufficient for almost complete localization.  However, in view
of the well-known difficulty to incorporate impurities as electrically active dopants in AlGaAs 
dielectric mirrors with high Al content, it is possible that the effective electric field across 
the superlattice in sample A is significantly smaller than 50 kV/cm.  Indeed, if we interpret the
\(X_1 \) line as the +1 exciton, then the effective field is 15 kV/cm, based on the
\(X_0 \) - \(X_1 \) energy separation.

\begin{figure}
\psfig{file=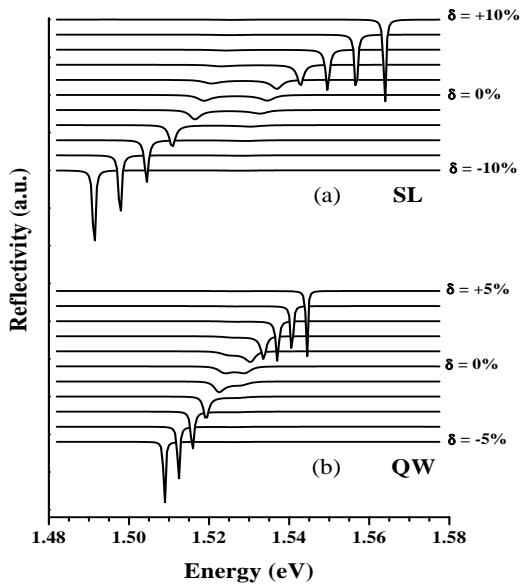,height=3.5in,width=3in}
\caption{Calculated reflectivity for a superlattice-cavity (a) and a quantum well-cavity (b) system.  In the former, 30 GaAs absorbers, each 80 \({\rm {\AA}}\) thick and separated from each other by 20 \({\rm {\AA}}\) of \({\rm Al}_{{\rm .25}} {\rm Ga}_{{\rm .75}} {\rm As}\), fill the cavity.  In the latter, three 80 \({\rm {\AA}}\) GaAs absorbers are placed at the antinodes of the \(\frac{{3\lambda }}{2}\) cavity.  The parameter \(\delta \) represents the relative detuning between the pure exciton and cavity modes.  In (a) a minimum splitting of 15 meV between the two modes is obtained when \(\delta  = 0 \).  In contrast, for (b) the corresponding splitting is only 4.5 meV.}
\end{figure}

In spite of the difference between these two field values, we favor the interpretation for
\(X_1 \) in terms of spatially indirect excitons since it is consistent with the absence of the
\(X_1 \) line in samples B and C.  A light-hole exciton line would be present in the spectra of all
three samples.  The T-dependence PL spectra of sample A also favor that interpretation.  In addition
to a feature analogous to \(X_1 \), we resolved at low T a weak feature about 14 meV below that of
\(X_0 \), which eventually had a small anticrossing with the cavity-mode peak.  That feature can be
seen as the -1 exciton of the Wannier-Stark ladder.  The smaller anticrossing is then the result of 
the smaller binding energy of the \( \pm \)1 exciton, in comparison with that of the \(X_0 \) exciton.

Our main observation--an enhanced Rabi-splitting in sample A between the spatially direct exciton
and the cavity mode--agrees with a simple classical simulation of the reflectivity for a set of 30
GaAs absorbers (each 84 \({\rm {\AA}}\) thick and 20 \({\rm {\AA}}\) apart from each other) 
distributed throughout the cavity defined by two dielectric mirrors.  Figure 4 shows the calculated
reflectivity spectra in the strong-coupling regime, for various values of the detuning parameter
\(\delta \), defined as \(\delta  = 1 - \frac{{\nu _{exc} }}{{\nu _{cav} }}\).  For simplicity, we 
have kept constant the exciton energy h\(\nu _{exc} \) at 1.5098 eV and varied the characteristic 
energy of the cavity, h\(\nu _{cav} \).  We have used "realistic" values for the linewidth and 
oscillator strength of the exciton, 4 meV and 0.035 respectively, so that the calculated lineshapes
resemble the experimental curves at 238 K. With this set of parameters the anticrossing is about
15 meV.  In contrast, a similar simulation for a microcavity with only three 84 \({\rm {\AA}}\)-thick
GaAs wells, each located at one of the antinodes of a \(\frac{{3\lambda }}{2}\) cavity, yielded
a minimum separation three times smaller.  The large difference in splitting between the two 
configurations confirms qualitatively our observations.

In this work, we have taken advantage of the built-in electric field of a p-i-n junction to
illustrate how, using a simple scheme, an electric-field-localized superlattice embedded in a
microcavity can enhance the exciton-photon coupling.  Further enhancements are expected by 
applying a larger (external) electric field or by optimizing the well-to-barrier thickness ratio
so that complete exciton localization is achieved.  The ideas presented in this paper can also be 
combined with existing schemes (e.g. extending the superlattice into the mirrors) and/or implemented 
in other semiconductor materials (e.g., II-VI compounds) to obtain even larger Rabi splittings.

This work has been sponsored in part by the Army Research Office and by the Spanish Interministerial
Commission for Science and Technology (CICyT).  We acknowledge partial support from a Turner Fellowship
(J. H. D.) and a Spanish CAM fellowship (S. M.).


\begin{references}
\bibitem{Zhu:90} Y. Zhu, D. J. Gauthier, S. E. Morin, Q. Wu, H. J. Carmichael, and T. W. Mossberg, Phys. Rev. Lett. {\bf 64}, 2499 (1990).
\bibitem{Weisbuch:92} C. Weisbuch, M. Nishioka, A. Ishikawa, and Y. Arakawa, Phys. Rev. Lett. {\bf 69}, 3314 (1992).
\bibitem{Savona:95} V. Savona, L.C. Andreani, P. Schwendimann, and A. Quattropani, Solid State Commun. {\bf 93}, 733 (1995).
\bibitem{Stanley:96} R. P. Stanley, R. Houdr\'{e}, C. Weisbuch, U. Osterle, and M. Ilegems, Phys. Rev. B {\bf 53} 10995 (1996).
\bibitem{Houdre:94}  R. Houdr\'{e}, C. Weisbuch, R. P. Stanley, U. Oesterle, P. Pellandini, and M. Ilegems, Phys. Rev. Lett. 73, 2043 (1994).
\bibitem{Tao:98} I. W. Tao, J. K. Son, C. Pecharrom\'{a}n, E. E. Mendez, and R. Ruf, Physica E 2, 685 (1998).
\bibitem{Pratt:98} A. R. Pratt, T. Takamori, and T. Kamijoh, Phys. Rev. B. 58, 9656 (1998).
\bibitem{Houdre:94b} R. Houdr\'{e}, R. P. Stanley, U. Oesterle, M. Ilegems, and C. Weisbuch, Phys. Rev. B. 49, 23 (1994).
\bibitem{Dang:98} L. S. Dang, D. Heger, R. Andr\'{e}, F. Boeuf, and R. Romestain, Phys. Rev. Lett. {\bf 81} 3920 (1998).
\bibitem{Boeuf:00} F. Boeuf, R. Andr\'{e}, R. Romestain, L.  Dang, E. Peronne, J. F. Lampin, D. Hulin, A. Alexandrou, Phys. Stat. Sol. A 178, 129 (2000).
\bibitem{Nelson:96} T. R. Nelson, et al., Appl. Phys. Lett. {\bf 69}, 3031 (1996).
\bibitem{Pratt:99} A. R. Pratt, T. Takamori, and T. Kamijoh, Appl. Phys. Lett. {\bf58}, 1869 (1999).
\bibitem{Bloch:98} J. Bloch, T. Freixanet, J. Y. Marzin, V. Theirry-Mieg, and R. Planel, Appl. Phys. Lett. {\bf 73} 1694 (1998).
\bibitem{Tredicucci:95} A. Tredicucci, Y. Chen, V. Pellegrini, and C. Deparis, Appl. Phys. Lett. {\bf 66} 2388 (1995).
\bibitem{Dickerson:01} J. H. Dickerson, et al., Bull. Am. Phys. Soc. {\bf 46}, 1228 (2001).
\bibitem{Wannier:60} G. H. Wannier, Phys. Rev. {\bf 117}, 432 (1960). 
\bibitem{Mendez:88} E. E. Mendez, F. Agull\'{o}-Rueda, and J.M. Hong, Phys. Rev. Lett. {\bf 60} 2426 (1988).
\bibitem{Info1} The difference between \(E1_{hh} \) and \(E1_{lh} \) is 16 meV.  \(m_{hh}  = .34m_e \).  \(m_{hh}  = .094m_e \).  Valence band barriers of 138 meV and 193 meV for x = 0.25 and 0.35, respectively (x is the Al mole fraction).
\end{references}
\end{document}